**Managing Uncertainty in Life Detection:**
**Venus and Mars Claims Compared**

Daliah Bibas[1] and Clément Vidal[2]
Centre Leo Apostel,
Vrije Universiteit Brussel,
Brussels, Belgium

**Abstract:**

Detection claims surrounding life beyond Earth often attract intense attention because the question of whether we are alone in the cosmos carries unusually high scientific, social, philosophical, and theological stakes. This chapter compares two high uncertainty life-detection claims, the 2020 phosphine detection claim on Venus and the 1976 Viking life-detection experiments on Mars, to examine how evidence, interpretation, and public narratives interact when results are ambiguous. We trace how each case unfolded through critique, reanalysis, and shifting assessments, and how institutional choices shaped subsequent research pathways, from long term monitoring and renewed Venus exploration to a Mars strategy centred on habitability rather than direct life-detection. Drawing lessons across both cases, we propose best practices for scientists, institutions, the wider scientific community, and science communicators, with emphasis on open science, constructive follow-up, cognitive bias awareness, and careful framing so that initial detections are not framed as final conclusions. We argue that managing future life-detections is less about eliminating uncertainty than about building transparent verification pathways that protect credibility as follow-up evidence refines the picture.

[1] Secondary Affiliation: SETI Post-Detection Hub, dalbibas@gmail.com
[2] Secondary Affiliation: Blue Marble Space Institute for Science, contact@clemvidal.com

## 1. Introduction

Claims about life beyond Earth have always drawn intense public attention. This is not a result of social media or modern news outlets, but instead due to the subject itself. The question of whether we are alone in the universe carries importance across cultures, religions, and worldviews given its scientific, philosophical, and theological stakes (Vidal 2015; Dick 2018). A confirmed discovery of extraterrestrial life would radically redefine humanity's place in the cosmos, causing a global paradigm shift. Because of this, even ambiguous hints of life can become meaningful and widely shared, long before the scientific community has time to test, replicate, and constrain what the data really supports. This can often create confusion, misplaced expectations, and put pressure on the researchers involved. Early interpretations can also solidify into the perceived solution, while later reanalysis can sometimes be treated as unorthodox rather than as the ordinary self-correcting process of scientific research.

This chapter uses the 2020 phosphine detection claim on Venus and the 1976 Viking missions life-detection experiments on Mars as cases of high uncertainty in astrobiology. We compare both cases to derive key lessons about managing uncertainty. We then propose best practices for scientists, scientific institutions, the scientific community, and science communicators that aim to improve scientific reliability, support public understanding, and prepare the community during the whole process of future astrobiology discoveries.

## 2. Phosphine on Venus

The prospect of finding life in the clouds of Venus has been secondary compared to the prospect of finding life on Mars. However, Morowitz & Sagan (1967) already envisioned the possibility almost six decades ago. In the last few decades, scientists have found extremophile organisms across Earth able to thrive in environments once thought to be completely uninhabitable, such as in conditions of extreme temperature, pressure, acidity, salinity, and radiation (Merino et al. 2019). These relatively new discoveries have expanded our imagination about where life might persist in the cosmos, including on Venus, where its surface conditions are hot enough to destroy most instruments quickly, the pressure is crushing, and the chemistry is highly corrosive.

Most planets are expected to have gradient conditions, instead of having one uniform environment. So, even though Venus' surface is lethally hot, temperature varies dramatically with altitude. Parts of the Venusian atmosphere have a temperature range that seems to be compatible with known terrestrial extremophiles. Similarly, while pressure is at 93 bars near the surface of the planet, there is a band in the clouds at around 50 km high where pressure is closer to Earth's 1 bar level (Pätzold et al. 2007). Here, droplets could hypothetically provide microenvironments and contain niches that are much less hostile than the surface (Limaye et al. 2018).

Still, less hostile than the surface is still not the same as hospitable. The cloud aerosols are dominated by sulphuric acid, water is scarce, and the chemistry is strongly oxidising, with additional stresses from UV radiation and limited nutrients. Hence, the key scientific questions are whether any biology could remain stable under these constraints and how to tell the difference between biology and unfamiliar chemistry. A detection of a biosignature could reflect the presence of life or it could reflect abiotic pathways we are not currently aware of.

### 2.1 Initial interpretation

In September of 2020, a research team led by astronomer Jane Greaves published a report presenting evidence for the presence of phosphine in the Venusian atmosphere (Greaves et al. 2020). While on Earth, phosphine is considered a biosignature as it is almost exclusively produced by biological activity (Sousa-Silva et al. 2020), Greaves and her team emphasised in their paper that an abiotic origin could not yet be excluded, including chemical pathways that remain unknown or poorly constrained.

### 2.2 Global response

At first, the story was covered with clear excitement by media outlets from all over the world (e.g. Sinha 2020; Stirone et al. 2020; Weule 2020), but when two independent groups reanalysed the James Clerk Maxwell Telescope (JCMT) and the Atacama Large Millimeter Array (ALMA) data, they pointed to problems with the ALMA baseline calibration (Snellen et al. 2020; Villanueva et al. 2020). The controversy was then rapidly amplified in the global press, with many reporting that the original claim had effectively been overturned (e.g. Domínguez 2020; Intini 2020; Siegel 2020). Within a matter of weeks, the original authors, working with the support of the North American and ESO ALMA Regional Support Centres, located and corrected the issues with the observations and processing pipeline (Bibas and Vidal 2025). After the ALMA data release concerns were addressed, the concluding signal strength dropped relative to the initial report, yet phosphine was still reported as present (Greaves et al. 2021). Some research groups remained unconvinced, suggesting that the signal could have been misinterpreted, potentially as a result of contamination with sulphur dioxide or instrumental noise (Thompson 2020; Akins et al. 2021; Lincowski et al. 2021; Villanueva et al. 2021).

### 2.3 Current standing

Although the case is still open, the overall direction of opinion among scientists has drifted toward scepticism while only a small minority of researchers remain cautiously optimistic. Many now argue that the existing results are not sufficient enough for the community to treat phosphine as a broadly accepted detection, let alone as a credible biosignature for Venus, without substantially stronger independent confirmation (for a synthesis of the debate, see Clements 2023). Subsequent work has also shifted from independent reanalyses to a structured follow-up, with long-term monitoring designed to measure phosphine alongside key confounders in the same observing setup (Clements 2024).

The debate and uncertainty have one important positive outcome, namely to motivate further study and exploration. As a matter of fact, NASA has two major upcoming missions towards Venus. Planned for mid-2031, the Deep Atmosphere Venus Investigation of Noble gases, Chemistry, and Imaging (DAVINCI) mission will send an orbiter and a probe that will descend through the atmosphere while analysing it (Garvin et al. 2022). The second mission also planned for 2031 is Venus Emissivity, Radio Science, InSAR, Topography, and Spectroscopy (VERITAS) that will map the surface of Venus at a high resolution using various instruments (Smrekar et al. 2022). In collaboration with NASA, the ESA also has a mission planned for 2031, called EnVision, that will do high-resolution radar mapping and atmospheric studies (Rosenblatt et al. 2021).

In addition to these, the Venus Life Finder mission is a privately-funded mission planned for summer 2026 and designed more explicitly to search for organic compounds within the planet’s atmosphere through a small (40 cm) titanium atmospheric probe (French et al. 2022). Remarkably, it is the first privately funded mission to another planet.

The controversy of the phosphine detection highlights a maturing astrobiology community that is increasingly learning how to carefully handle ambiguity without rushing to closure. Given that evidence of life on Venus would be consequential for space science, but also for society and human psychology (Vidal and Bibas 2025), continued engagement from the scientific community is not just expected, it is needed. Ultimately, in this case, what helped the field move towards a more cautious, evidence-based evaluation of the phosphine detection was the combination of independent reanalysis, support from observatory support centres, and the willingness of the original authors to re-examine their methods and interpretations.

## 3. Viking Missions on Mars

The prospect of finding life on Mars has long been at the forefront of the field of astrobiology, as it has long been considered a very plausible target to find life in our solar system. It is nearby, accessible compared to Venus, and has similar geological features to Earth. There is strong evidence that Mars once hosted large amounts of liquid water and that certain past environments were habitable (Ehlmann and Edwards 2014; Grotzinger et al. 2014). Today, the Martian surface is cold, dry, oxidising, and exposed to high levels of radiation. However, several lines of research suggest that Mars may still offer localised habitable niches, most plausibly below the surface where rock and ice can shield from radiation and where small amounts of liquid water could exist transiently (Dartnell et al. 2007; Martínez and Renno 2013).

### 3.1 Findings

In 1976, the NASA Viking 1 and 2 landers arrived on the surface of Mars and carried out the first direct life-detection experiments on another planet. Each lander conducted three biological experiments: the Labeled Release (LR), which looked for gases that could be produced by microbial metabolism; the Gas Exchange (GEX), which monitored how the gas composition above soil samples changed; and the Pyrolytic Release (PR), which searched for carbon fixation processes similar to photosynthesis (Klein et al. 1976). Additionally, each lander possessed a Gas Chromatograph-Mass Spectrometer (GCMS), a chemical analysis instrument used to investigate the soil and search for organic molecules. The LR experiment, developed by Levin and Straat (1977), added a nutrient solution marked with radioactive

carbon-14 to the soil and monitored any gases produced. The rapid production of carbon dioxide, observed by both Viking landers thousands of kilometres apart, was initially interpreted as a possible sign of microbial activity. Yet the GCMS results showed no sign of organic molecules in the same soil, and those of the GEX and PR experiments were interpreted as either ambiguous or negative (Klein et al. 1976). These conflicting results led to NASA emphasising in press conferences and official reports that the Viking landers had not provided definitive evidence of life on Mars (French 1977; Klein 1977).

### 3.2 Global response

While the potential signs of life were highlighted by media outlets, the coverage still stressed how the experimental results remained unresolved (e.g. McElheny 1976; Cooper 1979). Decades of debates followed. Many researchers insisted that the lack of detected organics, together with evidence for strong oxidising agents in the soil, pointed toward a non-biological explanation (e.g. Mazur et al. 1978; Klein 1999). At the same time, other researchers continued to publish work in support of a biological interpretation (e.g. Aksyonov 1979; Levin and Straat 1981), with more recent work noting that highly reactive chlorine-based salts called perchlorates, only discovered on Mars decades later, could have broken down organic molecules during the GCMS heating, accounting for the negative results (e.g. Houtkooper and Schulze-Makuch 2010; Navarro-González et al. 2010).

The institutional response from NASA was to stop further funding for life-detection missions on Mars. As Gilbert Levin reports, he and Patricia Straat submitted several proposals over the years but “were always rejected with the flat statement that no life-detection tests would be funded” (Warmflash 2016). Maybe this decision is not that surprising. Indeed, after having spent one billion dollars to tackle the question “Is there life on Mars?”, NASA was unlikely to anticipate obtaining ambiguous results that are still open to interpretation as of today. This outcome may have been seen as a failure, so choosing to stop further funding direct life-detection efforts could have seemed most logical.

Instead of focusing on a binary result that would give a yes/no answer to the question of "Is there life?”, further missions were developed around the concept of *habitability*, which can be qualified not as a binary choice, but as a gradient. By defining missions in terms of evaluating habitability, NASA can ensure that there is a sustained flow of scientific findings to strengthen the rationale for the investment.

### 3.3 Current standing

For the next fifty years, most researchers studying Mars considered the Viking results as best explained by non-biological processes, although a small group still interpreted the LR data as evidence for living Martian organisms (e.g. Bianciardi et al. 2012; Schulze-Makuch 2024). Today, soil analysis from more recent missions to Mars, including the detection of complex organic molecules by NASA's Curiosity rover in 2018 (Eigenbrode et al. 2018; Freissinet et al. 2025), is prompting some researchers to reconsider their initial interpretations of the Viking results.

## 4. Lessons Learned and Best Practices

While the two cases present two different types of detected signatures (see Table 1), the Mars Viking missions being an in-situ detection of carbon dioxide release from the soil and the Venus phosphine case being a remote detection of phosphine in the planet's atmosphere, both cases moved quickly from detection to broad public attention while key uncertainties remained unresolved. Most experts currently lean toward abiotic explanations in both the Viking and phosphine cases, however, neither case has fully closed. Correction processes are still unfolding. The remaining uncertainties have led to radically different outcomes: there has not yet been another life-detection mission on Mars since the Viking missions, while there are many missions planned to investigate the atmosphere of Venus.

| | | Case | |
|---|---|---|---|
| | | Viking Landers (1976) | Venus Phosphine (2020) |
| Feature | Signature | In-situ biosignature | Atmospheric biosignature |
| | Detection | Carbon dioxide release from soil | Phosphine gas in Venus' atmosphere |
| | Media Reception | Confusion | Excited to Dissapointed |
| | Scientific Debate | Interpretation questioned | Array calibration questioned |
| | Correction Processes | Ongoing | Ongoing |
| | Expert Consensus | No consensus; majority believe in non-biological explanation | No consensus; majority believe in non-biological explanation |
| | Time until Consensus | Ongoing debate for ~50 years | Ongoing debate for ~5 years |
| | Ethical Considerations | Results presented with caution, acknowledging uncertainties | Initial findings presented with caution, acknowledging uncertainties |

**Table 1.** Overview of key features of the Viking landers and Venus phosphine detection claims. Adapted from (Bibas and Vidal 2025).

Let us now propose some best practices for scientists, scientific institutions, the scientific community, and science communicators.

### 4.1 Scientists

The necessity of open science is critical for quick independent verification and progress (Woelfle et al. 2011). All data and analysis software are part of a given claim and should be made available. The phosphine debate clearly shows how independent scrutiny depends on access to data.

Responsible scientists avoid overinterpretation of ambiguous data and proactively engage with criticism as part of the process. This way, analyses can be refined, better evidence can be collected, and models can be strengthened. For instance, astronomer Jane Greaves and her team exemplified this by revising their analysis of phosphine on Venus when calibration errors were identified by other researchers.

On the psychological side, scientists participating in the discovery will get immediate worldwide attention. This inevitably includes the expression of all points of view and all attitudes, good and bad. Scientists need to be prepared psychologically for this avalanche that may include insults, harassment, ridicule, or any *ad hominem* attacks such as gender-based ones. Maintaining a professional reputation may prove extremely difficult during and after such a storm.

### 4.2 Scientific Institutions

Institutions must also recognise the personal toll of high-uncertainty claims. Since the search for life beyond Earth often triggers intense public reactions, institutions must proactively develop safety and support strategies to protect scientists (Bimm et al. 2025).

The mechanisms of power and control were different in the 1970s than they are in the 2020s. In the closely related domain of SETI, the upcoming post-detection protocols now reflect this completely different media landscape (Garrett et al. 2025). There was no internet and no social media during the Viking missions, so the data mostly stayed with NASA and the control of the Viking narrative was vertical: Gerald Soffen, the project scientist at NASA headquarters held the ultimate authority to officially conclude that life was not detected. But in the case of the phosphine debate, the community of scientists could quickly critique and self-correct their works via the publication of contradictory preprints, a quick dynamic that was simply impossible in the 1970s.

Institutions that operate shared facilities have responsibilities that go beyond allocating telescope time. When they sit at the centre of a high attention claim, it is critical that they provide resources to actively support follow-up troubleshooting, reanalysis, and verification, including the launch of new missions. For example, within a matter of weeks after the problems with the ALMA baseline calibration were pointed out, the North American and ESO ALMA Regional Support Centres worked with the original authors to locate and correct the issues, showing a high level of responsibility and a strong institutional backing.

As the community report from the biosignatures standards of evidence workshop (Meadows et al. 2022) recommends, national agencies and scientific institutions could set aside dedicated funding for life-detection verification costs.

### 4.3 Scientific Community

When a new detection of life is published, the first reaction of the scientific community is often to challenge the claim by deconstructing and debunking instead of constructing and proposing new tests or follow-up observations. But the evaluation of such claims is rarely settled by a single result. Instead, it emerges through the gradual integration of multiple findings (Meadows et al. 2022).

Furthermore, researchers may be influenced by an *anchoring bias*, placing too much weight on earlier interpretations. For example, more recent Mars missions have provided further information about Martian soil chemistry, and findings such as NASA Curiosity rover's 2018 report of complex organic molecules (Eigenbrode et al. 2018; Freissinet et al. 2025) have led some researchers to revisit how the Viking life-detection results should be interpreted. As a result, the scientific community may give less careful attention to alternative past experiments, whether consciously or not. Re-opening and re-evaluating past experiments is often a harder path than presenting clear, new observations or results.

### 4.4 Science Communicators

Science communicators, whether journalists, science writers, or public information officers, have an immense responsibility of creating a first impression that fixes the minds of millions, both scientists and the general public. This includes not only the ideas and arguments but also the choice of illustrations (accurate or inaccurate) that also fix the imagination. The first impression often sticks, which means communicators can unintentionally create anchoring effects that shape how the entire story is understood. Hence, science communicators should avoid treating early ideas as the final consensus and use explicit language to separate detection from interpretation. Because astrobiological claims have such a high visibility, communicators should be especially mindful that *ad hominem* or poorly researched statements can be highly detrimental to the professional careers of scientists.

The science communicator's goal is not only to relay information accurately but also to have their pieces read by a maximum of people. These two objectives are in tension. Putting too much weight on reaching a maximum audience leads to clickbait headlines and sensational language that can dissipate the sense of uncertainty at the core of the scientific process. Indeed, research confirms this risk, showing that news coverage consistently inflates the certainty of outcomes and the significance of evidence, effectively prioritising speculation over scientific nuance (Albergaria et al. 2025). On the other hand, a very careful, technical report conveying a maximum of the nuance and uncertainty of the original scientific work would be difficult to understand for a majority of readers. Science communication is the art of balancing this tension.

For readers of the news, the phosphine debate may have seemed like a roller-coaster, shifting from “Life on Venus? Astronomers See a Signal in Its Clouds” (Stirone et al. 2020) to “Venus is Dead: New Analysis Shows Phosphine, A Possible Biosignature, Is Absent.” (Siegel 2020) in only 39 days time! As we proposed earlier (Bibas and Vidal 2025), science communicators could aim to trigger a sense of mystery by spelling out the scientific uncertainties, and try to portray the scientific debates as a suspenseful story instead of taking a side and claiming a definitive resolution.

## 5. Conclusion

The search for life beyond Earth comes with high stakes, not just scientifically, but also socially, philosophically, and theologically. A discovery will likely not arise from one flashy moment, but from a slow, methodical process that may be messy and defined by ambiguity. Uncertainty is psychologically inconvenient and uncomfortable, but scientifically ubiquitous. The main challenge is not to eliminate all uncertainty, as that would not be realistic, but instead to manage it well. In astrobiology, the sources of uncertainties are numerous due to the remote locations of the targets, the extreme environments, and the difficulty to design missions that would provide unambiguous results. In addition, astrobiologists have institutional and social pressure to have a clear answer of whether or not life exists elsewhere. After all, this typically drives their motivation and passion!

In our case studies, the reduction of uncertainty was deployed by two very different strategies. The ambiguous results of the Viking experiments ended up cutting further life-detection efforts on Mars and shifting focus to habitability, which can be assessed as a gradient and not as a binary outcome. The dynamic is the opposite for Venus: the debate over phosphine has reignited interest, directly motivating upcoming missions such as DAVINCI, VERITAS, EnVision, and the Venus Life Finder.

To navigate future claims responsibly, a coordinated effort is required across the scientific process. *Scientists* must comply with open science, be willing to re-open past cases, and prepare psychologically for intense public scrutiny. *Institutions* must provide the resources necessary for immediate follow-up and reanalysis. Simultaneously, the *scientific community* must guard against cognitive biases and provide not only sceptical reactions but also constructive criticism, and *science communicators* must exercise caution to ensure that initial detections are not framed as final conclusions.

By adopting these best practices for uncertainty management of life-detection claims, the astrobiology community can ensure that upon the next potential discovery, the process of verification will be as robust as the discovery is profound.